\documentclass[12pt]{article}
\usepackage{setspace}
\usepackage{overcite}
\usepackage{amssymb}
\usepackage{amsmath}
\usepackage[dvipdfm]{graphicx}
\usepackage{enumerate}
\usepackage[bf]{caption2}

\def\thebibliography#1{\section*{{\sf REFERENCES AND NOTES}}\list
 {\arabic{enumi}.}
 {\settowidth\labelwidth{[#1]}\leftmargin\labelwidth
 \advance\leftmargin\labelsep \usecounter{enumi}}
 \def\newblock{\hskip .11em plus .33em minus .07em}
  \parskip -0.7ex plus 0.5ex minus 0ex
 \sloppy\clubpenalty4000\widowpenalty4000
 \sfcode`\.=1000\relax}

\begin{document}

\noindent{\Large\textbf{Charge Carrier Transport in Disordered Polymers}}

\bigskip
\noindent\textbf{S. V. Novikov}

\noindent A.N. Frumkin Institute of Electrochemistry, Leninsky prosp. 31, Moscow 119071, Russia

\begin{abstract}
\noindent
General properties of charge carrier transport in disordered
organic materials are discussed. Spatial correlation between
energies of transport sites determines the form of the drift mobility field
dependence. Particular kind of spatial correlation in a disordered
material depends on its nature. Mobility field
dependences have to be different in polar and nonpolar materials.
Different methods of mobility calculation from the shape of
photocurrent transient are analyzed. A widely used method is very
 sensitive to the variation of the shape of the transient and
 sometimes produces results that effectively masquerade the true
 dependence of the mobility on electric field or trap
 concentration. Arguments in favor of the better, more reliable method are suggested.
Charge transport in materials containing charged traps is
considered without using the isolated trap approximation and this
leads to qualitatively different results. They indicate that
the effect of charged traps can hardly be responsible for
experimentally observed transport properties of
disordered organic materials.

\noindent
\textbf{Keywords:} charge transport; correlated disorder;
computer simulations
\end{abstract}


\section*{{\sf INTRODUCTION}}

Any feasible electronic application of a polymer material implies
that the material can conduct an electric current via some kind of
charge transport mechanism. Velocity of charge carriers (electrons
or holes) may be an important parameter per se effectively
determining operating characteristics of the device or it could
influence other important parameters such as carrier injection
or recombination efficiency. Some degree of structural disorder is present in
most polymers currently used in real life applications, so they
naturally belong to the general category of disordered organic materials.
At the same time disordered low molecular weight organic materials
often demonstrate similar general features of charge transport;
for this reason in this article various properties of charge transport in
disordered organic materials are mostly discussed without specific emphasis on
polymeric nature of a material.

For many years the study of charge carrier transport in disordered organic materials
was mostly motivated by hope to use them in various electronic and optoelectronic devices.
 At the same time this problem is an interesting and
sometimes even intriguing area of modern condensed matter physics.
Good review of the area may be found in recent monographs
\cite{pope,bw_book}. We consider here only materials which are insulators
under usual conditions and attain short-lived conductivity after
injection of charge carriers under the action of laser pulse or electric
discharge and completely leave aside such highly conducting materials as
doped polyacetylene, polyanyline etc. Typical materials include low molecular
weight organic glasses, polymers doped with aromatic organic molecules
(usually amines or hydrozones), polysilanes, polyconjugated
polymers, and other compounds. This diverse variety demonstrates
a lot of common transport features unambiguously indicating a common
transport mechanism.

Transport properties of materials are usually characterized by
carrier drift mobility $\mu$ and its dependence on applied electric
field $E$, temperature $T$, and other relevant parameters. The most
direct method of the mobility measurement is the
time-of-flight method (TOF) where carriers are generated at the
vicinity of one electrode of the plane capacitor (organic
material fills the space between electrodes) and then drift to
the opposite electrode under the action of applied uniform
electric field $E$. Analyzing the shape of the
photocurrent transient, one can calculate the transit time
$t_{\rm tr}$ (the moment when current begins to drop, thus indicating
 arrival of carriers to the opposite electrode) and then the mobility
\begin{equation}
\mu=\frac{v}{E}\approx\frac{L^2}{V_0 t_{\rm tr}},
\label{mu}
\end{equation}
where $L$ is thickness of the transport layer, $v$ is the
average velocity of charge carriers, and $V_0$ is the applied
voltage. Dominant majority of experimental data cited in
this article was obtained by the TOF method.

Let us briefly summarize major experimental findings offering important clues
on the nature of charge transport in such materials.\cite{pope,bw_book,tapc,bassler}
\begin{enumerate}[1)]
\item Molecularly-doped polymers offer a natural possibility to study dependence
of the mobility on dopant concentration.
It was found that
\begin{equation}
\ln\mu \approx -2R/R_0,
\label{mu(R)}
\end{equation}
where $R$ is the average distance between dopant molecules. This
relation suggests that the transport occurs as a series of hops
between localized states (transport sites) originating, in
this particular case, at the dopant molecules. However, in some exceptional cases
the radius $R_0$ of the wave function of a transport
site is unreasonably large \cite{polysterene}.

\item Mobility increases exponentially with the increase of temperature.
There is no consensus on the form of this dependence:
some authors favor a simple activation dependence
\begin{equation}
\ln\mu\propto -1/T,
\label{mu(T)_1}
\end{equation}
while others argue that the dependence
\begin{equation}
\ln\mu\propto -1/T^2
\label{mu(T)_2}
\end{equation}
is more accurate. It is generally accepted that an exponentially strong
temperature dependence may be most naturally explained by suggesting that
the major factor governing
charge transport in disordered organic materials is an energetic
disorder: random locations and orientations of
molecules in the bulk of the material produce random fluctuations of site energies $U_i$.
Then the dependence $\mu(T)$
should be connected with the density of states $P(U)$.
There is no reason for $P(U)$ to maintain the same form
in different materials and this may be the reason for different
forms of $\mu(T)$.

\item The most surprising fact is the mobility field dependence.
Indeed, for carrier hopping between spatially localized sites we should
expect
\begin{equation}
\ln\mu\propto eER/k_{\rm B}T,
\label{mu(E)_naive}
\end{equation}
just estimating a typical drop in the difference of carrier energy
between two transport sites separated by the typical intersite distance $R$.
Yet instead of eq \ref{mu(E)_naive} an almost ubiquitous
Poole-Frenkel (PF) dependence emerges in the materials
\begin{equation}
\ln\mu\propto \sqrt{E},
\label{mu(E)_PF}
\end{equation}
leading to much stronger field dependence of the mobility in
weak field. Uncertainty in the form of dependence $\mu(T)$ is mostly
originated from the difficulty to measure the
quasi-equilibrium (nondispersive) mobility in a wide temperature
range. In striking contrast to this limitation, dependence (\ref{mu(E)_PF})
was measured for some materials in really wide field range (an unrivaled
example is ref. \citen{spg} with $E$ varying from $8\times
10^3$ to $2\times 10^6$ V/cm).
\end{enumerate}

\noindent
For a long time an explanation of the PF dependence has been considered
as a major problem for the charge transport theory. Since the early paper of Gill,
who suggested an empirical formula
\begin{equation}
\mu=\mu_0\exp\left[\gamma\left(\frac{1}{T}-\frac{1}{T_0}\right)
\left(\sqrt{E}-\sqrt{E_0}\right)\right],
\label{Gill}
\end{equation}
based on the study of charge carrier transport in polyvinylcarbazole,\cite{gill}
it was very tempting to attribute the field dependence of $\mu$ to
the influence of charged traps.
In the case of trap-controlled charge transport with charged particles
(having charge of the opposite sign to carriers) serving as traps, an
applied electric field $E$ leads to the decrease of activation energy
of the carrier escape from a trap. In this case
\begin{equation}
\gamma=\frac{2}{k_{\rm B}}\left(\frac{e^3}{\varepsilon}\right)^{1/2}
\label{PF}
\end{equation}
(here $\varepsilon$ is a dielectric constant)
and calculated values of $\gamma$ are usually close to the
measured ones by the order of magnitude, though discrepancies
by a factor of 2--3 are typical. Yet this classical PF mechanism
attained a lot of criticism because there is no evidence for
charged traps to be a common constituent in very different
classes of disordered organic materials \cite{spg}.

A first really successful  transport model was
the Gaussian Disorder Model (GDM) \cite{bassler}.
In the GDM transport sites are arranged on a regular lattice and
are assigned site energies $U_i$ drawn
independently from a Gaussian distribution having variance $\sigma^2$.
This model explained many features of charge transport in
disordered organic materials, among them the temperature dependence
of mobility, transition from the quasi-equilibrium transport to
the dispersive-like with temperature decrease, and others features.
Still the explanation of the PF field dependence
remained an unsolved problem because the GDM can reproduce
this dependence in the limited field range only, not significantly
wider than $3\times 10^5 - 1\times 10^6$ V/cm, in weaker field dependence
$\mu(E)$ has the usual form (\ref{mu(E)_naive}). Moreover, the only reason
for the approximate PF dependence in the GDM is the Miller-Abrahams
hopping rate \cite{prl}. There is no reliable evidence that this particular
rate is the universal hopping rate in a wide variety of disordered materials.
For this reason it was highly desirable to retain all useful features
of the GDM and yet incorporate some new feature permitting to explain
the PF field dependence.

\section*{{\sf CHARGE TRANSPORT IN POLAR MATERIALS}}

A history of the GDM modification that eventually lead to the
successful explanation of the PF dependence (\ref{mu(E)_PF}) began with the study
of dipolar contribution to the total energetic disorder
in organic materials. Computer simulation of the distribution of
energies for a carrier moving through a simple cubic lattice with sites
occupied by randomly oriented dipoles (thus, the energetic disorder
in this case originates from the electrostatic charge-dipole
interaction)
was carried out by Dieckmann, B\"{a}ssler, and Borsenberger \cite{dieckmann};
they found that $P(U)$ has the Gaussian form for high
concentration of dipoles and goes to the Lorentzian one for low
dipole concentration.
They calculated the energy variance $\sigma^2_{\rm d}=\left<U^2\right>$
and found that typical values of $\sigma_{\rm d}$ are
0.05--0.1 eV which means that the dipolar contribution should be a
significant part of the total $\sigma$ (according to the GDM, \cite{bassler}
$\sigma\backsimeq 0.1$ eV is needed to fit experimental temperature
dependence of $\mu$).  Later it was found that for
this model of dipolar glass (DG) function $P(U)$ could be calculated
analytically \cite{JETP}. For a simple cubic lattice
\begin{equation}
\sigma_{\rm d} = 2.35\frac{epc^{1/2}}{\varepsilon a^2}
\label{sd}
\end{equation}
(here $a$ is a lattice scale, $p$ is a dipole moment, and $c$
is a fraction of sites occupied by dipoles)
and the distribution of $U$, while indeed having the Gaussiam form
for $c\backsimeq 1$, has the long tail $P(U)\propto 1/U^{5/2}$ for
$c \ll 1$.
\begin{figure}[htb]
\begin{center}
\includegraphics[width=3in]{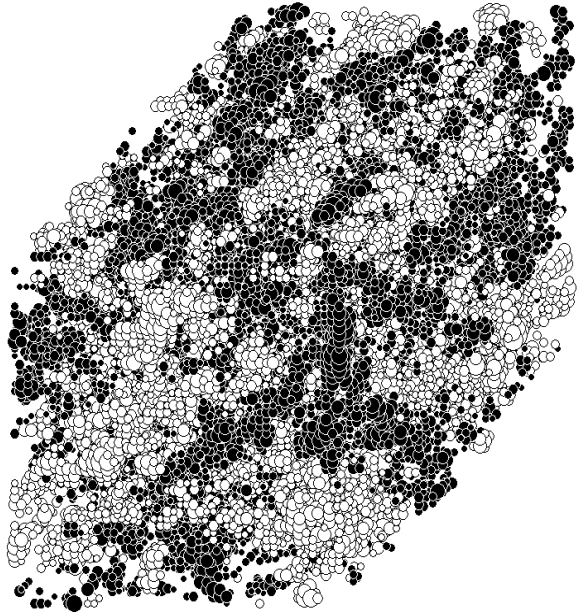}
\end{center}
\caption{Distribution of site energies $U_i$ in $50\times 50\times 50$ sample of the simple cubic lattice model of dipolar glass.
Black and white spheres represent sites with
positive and negative values of $U_i$, correspondingly, while the radius
of a sphere is proportional to the absolute value of $U_i$. Sites with
small absolute values of $U_i$ (less than $3ep/\varepsilon a^2$) are not
shown for the sake of clarity. }
\label{fig1}
\end{figure}

The most significant property of the DG model is a strong spatial
correlation in the distribution of $U(\vec{r})$ and the energy
correlation function decays very slowly with distance \cite{nv}
\begin{equation}
C(\vec{r}) = \left<U(\vec{r})U(0)\right> \approx 0.74\hskip2pt \sigma^2_{\rm d}\frac{a}{r},
\hskip10pt r \gg a
\label{c_dip}
\end{equation}
(here angular brackets denote statistical average).
In a correlated Gaussiam medium the conditional probability
$P(U|U_0)$ which is the probability density to have energy $U$ at
the site $\vec{r}$ if the site energy at the reference site $\vec{r}_0$ is $U_0$
has the form \cite{nv}
\begin{equation}
P(U|U_0)=\frac{1}{\sqrt{2\pi \sigma^2_\delta}}\exp\left[-\frac{1}{2\sigma^2_\delta}
\left(U-\frac{C(\vec{r}-\vec{r}_0)}{\sigma^2_{\rm d}}U_0\right)^2\right],
\hskip10pt \sigma^2_\delta=\sigma^2_{\rm d}-\frac{C^2(\vec{r}-\vec{r}_0)}{\sigma^2_{\rm d}}.
\label{cond}
\end{equation}
Keeping in mind that $C(0)=\sigma^2_{\rm d}$, we see that equation
\ref{cond} expresses the tendency of sites that are spatially close
to have close energies, in striking contrast to the GDM case (see Fig.\ref{fig1}).
\begin{figure}[htb]
\begin{center}
\includegraphics[width=3in]{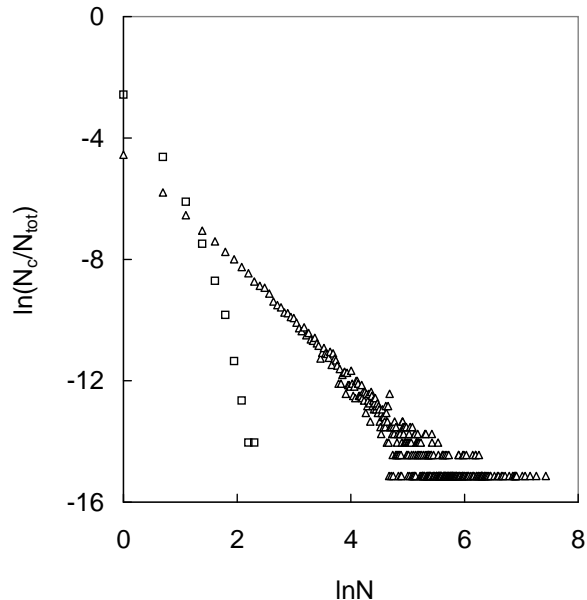}
\end{center}
\caption{Computer simulation of the cluster distribution on size (squares - the
GDM, triangles - DG model with same $\sigma$).
Here $N$ is the number of sites in a cluster, $N_c$ is the number
of clusters and $N_{\rm tot}$ is the total number of sites; cluster is defined as a connected set of sites having energies greater than some (arbitrary) threshold energy ($3.7\sigma$ in this particular case). }
\label{fig_c}
\end{figure}

Another manifestation of the correlated nature of DG is an exact
result for the dependence of the distribution of average domain energy
$U_V=\left<U\right>_V$ on domain size $V$.
For a Gaussian random medium without any
spatial correlation the distribution has the Gaussian form with
variance
\begin{equation}
\sigma^2_V=\sigma^2 a^3/V,
\label{domain_nc}
\end{equation}
while for spherical domains in DG \cite{pssb2000}
\begin{equation}
\sigma^2_V=12\sigma^2_{\rm d} a/5R,
\label{domain_dip}
\end{equation}
where $R$ is the radius of the domain. The difference between (\ref{domain_nc}) and
(\ref{domain_dip}) suggests that large domains are much more common
in dipolar matrix and, indeed, relative number of large clusters
in DG is greater by many orders of magnitude in comparison with
the uncorrelated medium (see Fig. \ref{fig_c}).\cite{icepom2}

\begin{figure}
\begin{center}
\includegraphics[width=3in]{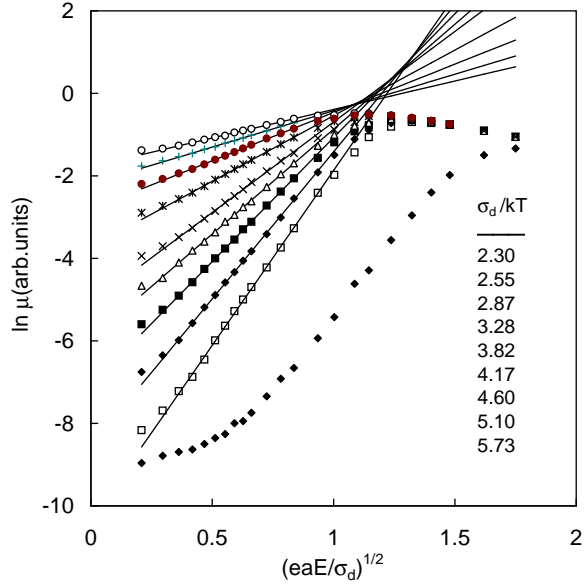}
\end{center}
\caption{Mobility field dependence in the DG model for different values
of $\sigma_d \beta$ (from top curve downward). The lowest curve is the
mobility for the GDM for $\sigma\beta=5.10$. If
 $\sigma_{d}=0.1$ eV and $a=10$\AA, then $eaE/\sigma_{d} \approx 1$
for $E=10^{6}$ V/cm.}
\label{3Ddip}
\end{figure}

At the same time Gartstein and Conwell suggested that a
correlated nature of the energy landscape should significantly
affect charge transport in disordered medium, effectively
increasing the typical scale of the hop and, thus, enhancing
mobility field dependence.  \cite{gc} The next major step was achieved by Dunlap,
Parris, and Kenkre, who found that for charge carrier
hopping in 1D random Gaussian energy landscape \cite{dpk}
\begin{equation}
\mu =\frac{\mu _{0}}{e\beta E \int\limits_{0}^{\infty }dy\exp \left(
-e\beta E y+\beta ^{2}\left[ C(0)-C(y)\right] \right) },
\hskip10pt \beta =1/k_{\rm B}T.
\label{1D}
\end{equation}
For the case of strong dipolar disorder $\sigma_{\rm d}\beta \gg 1$ equation
\ref{1D} gives
\begin{equation}
\mu \approx \mu _{0}\exp \left[-(\sigma_{\rm d} \beta)^{2}+
2(\sigma_{\rm d} \beta)\sqrt{ea \beta E}\right],
\label{1Dres}
\end{equation}
and, in general, if $C(\vec{r})\propto 1/r^{p}$ then
$\ln\mu\propto E^{p/p+1}$. At last, computer simulation confirmed
that the dipolar correlation leads to the PF mobility field dependence
for 3D carrier transport \cite{nv1,nv2} (see Fig. \ref{3Ddip}) and
on the basis of extended simulation an empirical relation was suggested \cite{prl}
\begin{equation}
\mu \approx \mu _{0}\exp \left[-\frac{9}{25}(\sigma_{\rm d} \beta)^{2}+
0.78\left[(\sigma_{\rm d} \beta)^{3/2}-2\right]\sqrt{eaE/\sigma_{\rm d}}\right].
\label{3Dres}
\end{equation}
Equations \ref{1Dres} and  \ref{3Dres} have essentially the same dependence on
 $E$ and $T$  and only numerical coefficients are different. Thus, the
simplified 1D model could serve for a quick estimation of the
mobility field dependence in disordered materials.

Good qualitative description of charge transport in a correlated
medium is presented in ref. \citen{dkp_image}. The principal difference
between a correlated random medium and an uncorrelated one is that
in correlated medium deepest valleys of the energetic landscape
(having maximal escape time for a carrier at $E=0$)
are also the widest. For this reason carrier escape time for such valleys
decreases drastically for $E > 0$.
For every magnitude of $E$ a critical size $R_c(E)$ exists
such that valleys of this size are the most efficient for carrier
capture. Simple calculation gives $R_c \propto E^{-1/2}$
 and $R$ in eq \ref{mu(E)_naive} should be
substituted for $R_c(E)$. \cite{dkp_image}
This is the physical reason for the PF dependence.

An attracting feature of the DG model is absence of any
significant free parameter: the energy scale (i.e. $\sigma$)
may be calculated using (\ref{sd}) provided $p$, $a$, and $c$ are known.
Still, numerical coefficients in eq \ref{3Dres} are certainly not
constants (for example, they weakly depend on $R_0$ \cite{parris}),
so equation \ref{3Dres} should be considered rather as a guide
and not as a strict relation for $\mu(E,T)$ in any particular polar material.

Another important feature of the DG model is that, unlike the GDM, the
form of the mobility field dependence for moderate field does not depend on the
particular kind of hopping rate: for $ea\beta E/\sigma_{\rm d}
\lesssim1$ mobility dependence retains its PF form (even the slope
is the same \cite{prl}) for any nonpathological hopping rate.

It is worth to mention, though, that in (partially) orientationally
ordered dipolar matrices correlation properties of the
distribution of
$U_i$ may be very different. In such matrices
transport properties may have nothing in common with the usual PF
picture. \cite{local}

\section*{{\sf CHARGE TRANSPORT IN NONPOLAR MATERIALS}}

A vast collection of experimental data indicates that the model of dipolar glass
cannot serve as a universal explanation for the PF mobility field dependence
in disordered organic materials. If the reason for this dependence
is the spatial correlation of the dipolar type (\ref{c_dip}), then it is
absolutely impossible to explain the existence of the PF dependence in
nonpolar materials, where it was routinely observed \cite{bgm,bgm2,tapc,ena,ttb}.
More close study of experimental data reveals, however, that in such materials the
experimentally tested field range is not too broad (not significantly greater
than one order of magnitude or even less) and, moreover, sometimes
clearly visible deviations from straight lines can be observed when
$\ln \mu$ is plotted against $E^{1/2}$ \cite{ttb}. In some cases these
deviations lead to the upward convexity of the mobility curve, while in
others they lead to the downward convexity. Quite formally, this behavior
may be described by
\begin{equation}
\ln \mu \propto E^n
\label{gener}
\end{equation}
with $n$ being in some cases smaller than 0.5 and in some cases
greater than 0.5. This observation hints that, quite possibly, in
weakly polar organic materials the real mobility field dependence
is not a true PF dependence, but rather a quasi-PF dependence with $n
\neq 0.5$. This quasi-PF dependence can successfully imitate true
PF dependence in not so wide field range. According to the result
of 1D model \cite{dpk}, the necessary condition for validity of
relation (\ref{gener}) is the algebraic behavior of correlation
function $C(\vec{r}) \propto r^{-p}$.

\subsection*{{\sf Quadrupolar Glass Model}}

One possible reason for such behavior is the contribution from
randomly oriented quadrupoles. Quite frequently, transport
molecules contain highly polar groups having significant dipole
moments. These groups may be arranged in such a way that the total
dipole moment of the molecule is close to zero, but the quadrupole
moment of the molecule is large enough. For example, for the
particular case of two identical groups having dipole moment $p$,
oriented in opposite directions and separated by distance $d$, the
total dipole moment of the molecule is exactly zero, but the
quadrupole moment $Q=pd$ is nonzero. In close analogy with the DG model
\cite{nv} we may consider the model of quadrupolar glass (QG),
its simplest incarnation being a simple cubic lattice with sites
occupied by randomly oriented quadrupoles \cite{QG_98}.

Calculation of the correlation function $C(\vec{r})$ for the QG model gives
\cite{QG_98}
\begin{equation}
C(\vec{r})\approx 0.5\sigma _{\rm q}^{2} \left( \frac{a}{r} \right)
^{3}, \hskip10pt r\gg a,
\hskip10pt \sigma _{\rm q}^{2} =C(0)=\frac{4e^{2} Q^{2} c}{5\varepsilon
^{2} } \sum\limits_{m}\frac{1}{r_{m}^{6} },
\label{eq10}
\end{equation}
where $c$ is the fraction of lattice sites, occupied by
quadrupoles. If $p = 3$D,
$\varepsilon = 3$, $a =10$ \AA, $b = 5$ \AA, then for the totally
filled lattice $\sigma_{\rm q} \approx 0.08$ eV. In 1D approximation the
carrier mobility in the QG model for the case of strong disorder
$(\sigma_{\rm q}\beta)^{2} \gg 1$ has the form
\begin{equation}
\mu \propto \exp \left[ -(\sigma_{\rm q} \beta)^{2} +\frac{2^{3/2}
}{3^{1/4} } (ea\beta E)^{3/4} (\sigma_{\rm q}\beta)^{1/2} \right]
\label{eq11}
\end{equation}
This particular field dependence can successfully imitate the PF
dependence in not too wide field range $E_{\rm max} /E_{\rm min}
\lesssim  10$. Moreover, there is another factor that can bring the
quadrupole field dependence even closer to the PF dependence. This
factor is an effect of dispersive transport: in dispersive regime
$\mu\propto  E^\alpha$.  \cite{sm} If plotted as $\ln \mu$  vs $E^{3/4}$, the
dispersive component bends the straight line, making the mobility
curve convex upward, thus pushing field dependence more close to
the PF one (contribution of the dispersive component is evident,
for example, for photocurrent transient in the data of ref. \citen{bgm2}
even at high temperature).

Detailed computer simulation of the 3D charge transport in the QG model
has not been carried out yet, but preliminary data unambiguously support the
principal result of 1D model for the mobility field dependence
$\ln\mu\propto E^{3/4}$ \cite{QG_2001}. Analogy with the DG model
suggests that the functional type of the mobility dependence on $E$ and $T$
could be well captured by the 1D model, while numerical coefficients should be
different. Extensive comparison of the QG model with experimental data is still absent,
though preliminary analysis of transport data for nonpolar polysilanes indicates
that dependence $\ln \mu =a+bE^{3/4}$ is certainly not worse (in terms of statistical
correlation coefficient $\mathcal{R}^2$) than the PF dependence \cite{nespurek}.

\subsection*{{\sf Los Alamos Model}}

Recently a new model was suggested to explain the PF mobility
field dependence in nonpolar polyconjugated polymers such as
polyphenylene vinylens (PPVs) \cite{LA1,LA2}. According to this
model, the major source of the energetic disorder in PPVs is
almost static intramolecular fluctuations of the torsion angle of
benzene ring, resulting in the carrier energy fluctuations
with the correlation function
\begin{equation}
C(\vec{r})\propto \frac{k_{\rm B}T}{r}\exp(-\alpha r), \hskip10pt
\alpha=s/K,
\label{LA}
\end{equation}
where $s$ is the intramolecular restoring force constant and $K$ is
the intermolecular restoring force constant. This function in the
limit case $\alpha \rightarrow 0$ gives an appropriate
behavior needed to provide the PF dependence. In densely packed PPVs
typical values of $\alpha$ should be
small, \cite{LA1,LA2} and according to the 1D model
the mobility field dependence has the form
\begin{equation}
\ln \mu \propto -\sigma_{\rm t}^2\beta+
\beta\sqrt{2\pi \sigma_{\rm t}^2 eaE},
\hskip10pt \sigma^2_{\rm t}=\nu^2/2\pi K a,
\label{LA_mu}
\end{equation}
where $a$ is the short-range cutoff and $\nu$ is the linear
electron-vibration coupling.

Field and temperature dependence of the mobility in the LA model
is different (though not very different) from the prediction of the QG model.
Thus, a careful study of the dependence of $\mu$ on $E$ and $T$ may provide
an opportunity to discriminate between these models. At present, there is no
possibility to state which particular model better describes the
mobility field dependence in nonpolar materials. We should note,
though, that the model of quadrupolar glass does not relay on any
particular property of the nonpolar material (apart from its quadrupolar
nature). From this point of view it is preferable over LA
model that, obviously, could hardly be applied for the explanation of
the properties of various low molecular weight nonpolar organic glasses.
These materials demonstrate the same PF field dependence as
PPVs and can be perfectly well described by the QG model.
High intramolecular flexibility is a necessary prerequisite for
significant amplitude of the energetic disorder in the LA model, yet experimental data
unambiguously demonstrate that some nonpolar materials consisting of
very rigid molecules may still have significant energetic disorder
(a good example of the material with planar rigid nonpolar molecules having
$\sigma\approx 0.09$ eV is discussed in ref. \citen{bgm2}).

It is worth to add that the very observation that served as a first clue for
the LA model, the significant difference between slopes of the PF
dependence for two particular polyconjugated polymers - one is traditional MEH-PPV
\cite{MEH-PPV}
and the other one (having very small slope of the mobility curve) is
stiff-chain polyfluorene \cite{bradley} - may be reasonably well explained
using the QG model. Indeed, the local stiffness of the main polymer
chain should by itself decrease the qudrupolar disorder. In
addition, the molecule of polyfluorene contains only carbon and
hydrogen atoms, while the molecule of more flexible MEH-PPV contains
oxygen atoms as well - this means that the qudrupolar moment of
the monomer unit for MEH-PPV is greater. Hence, both
factors acts in the same direction, decreasing the quadrupolar
disorder in polyfluorene and weakening the mobility field dependence.

\begin{figure}[htb]
\begin{center}
\includegraphics[width=3in]{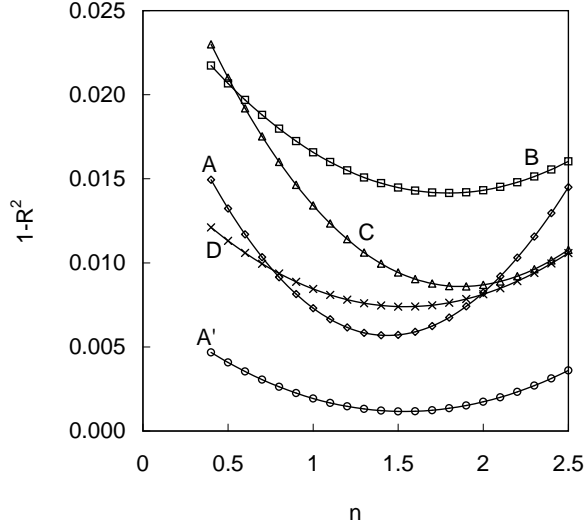}
\end{center}
\caption{The correlation coefficient $\mathcal{R}^2$ obtained
by fitting the temperature dependence of $b_{\rm PF}$ for different PPVs on $n$ in
$b_{\rm PF}\propto \beta^n$.
The best fit corresponds to the minimum of $1-\mathcal{R}^2$. Curves for different
PPVs from ref. \citen{blom2} are marked A, B, C, and D; curve A' represents data
from ref. \citen{blom1} (the same PPV as for the curve A, but different thickness of
the transport layer).}
\label{fig_fit}
\end{figure}

Recent papers on charge transport in PPVs offer a limited
opportunity to compare QG and Los Alamos models. \cite{blom1,blom2}
These papers offer data on the temperature dependence of the PF
coefficient $b_{\rm PF}=\gamma/T$ (see eq \ref{PF}). By definition
\begin{equation}
b_{\rm PF}=\frac{\partial \ln\mu}{\partial E^{1/2}}
\label{b_PF}
\end{equation}
and for the LA model $b_{\rm PF}\propto \beta$.
Although in the QG model coefficient $b_{\rm PF}$ is not a true constant
with respect to $E$, its field dependence is very weak
\begin{equation}
b^{\rm QG}_{\rm PF} \propto E^{1/4}\beta^{5/4}
\label{b_PF_QG}
\end{equation}
and we can approximately treat it as an effective constant. Results of
the fit $b_{\rm PF}\propto \beta^n$ are shown in Fig. \ref{fig_fit}.

For most PPVs the best $n$ is close to 1.5 and, thus, more close to 1.25
(QG model) than to 1 (LA model). A possible reason for the discrepancy with
the QG model may be the specific technique used in refs. \citen{blom1,blom2}
where mobility was calculated from the current-voltage characteristics
in the space-charge limited conduction regime assuming the PF mobility field dependence.
Thus, the mobility field dependence was not calculated at all, only
the coefficient $b_{\rm PF}$ was optimized to fit experimental current-voltage curves.
The direct study of the temperature
dependence of the corresponding true QG coefficient
\begin{equation}
b_{\rm q}=\frac{\partial \ln\mu}{\partial E^{3/4}}
\label{b_QG}
\end{equation}
obtained from TOF experiments should be a better way to test the QG model.

\subsection*{{\sf Mixed Disorder: Is It Possible to Provide a Simple Formula for
the Mobility Field Dependence?}}

In a typical case the total energetic disorder in a material has several
contributions of different nature
\begin{equation}
U_{\rm tot}(\vec{r})=\sum_s U_s(\vec{r})
\label{Utot}
\end{equation}
(say, dipolar and quadrupolar components).
In such situation the correlation function (assuming statistical
independence of individual components) is a sum of terms having
different dependence on distance. The resulting
mobility dependence in the 1D model has the form
\begin{equation}
1/\mu \propto \int_0^\infty dy \exp\left(-eE\beta y+\beta^2
\sum_s \left[C_s(0)-C_s(y)\right]\right),
\label{mu_tot}
\end{equation}
which cannot be written in the form of an explicit simple formula for nontrivial
functions $C_s(y)$.

Moreover, mobility field dependence could present a clear indication of
the existence of several contributions to (\ref{Utot}) only when mobility
is measured in an exceptionally wide field range.
Careful analysis shows that calculation of individual dipolar
and quadrupolar contributions to the total disorder could be
performed only in the case when mobility field dependence is measured
in the field range
spanning at least three orders of magnitude \cite{NIP15}. This is hardly
possible in real experiments.

\section*{{\sf TRAP-CONTROLLED CHARGE TRANSPORT}}

Most organic materials contain impurities serving as traps for
charge carriers (i.e. such impurities have
energy levels laying deeper than levels of majority of transport sites).
Typically, energies of trap sites have no spatial correlation and
produce an additional spatially uncorrelated disorder.
Important question is: to what extent general properties of charge
transport in correlated medium are insensitive to the presence of traps
(later we will limit our consideration to the the particular case
of dipolar medium)? This problem has been extensively studied in
the 1D approximation and by means of 3D computer simulation \cite{QG_98}.
Again, major results are the same for
both approaches and may be summarized as follows:
\begin{enumerate}[1)]
\item addition of traps leads to the decrease of the carrier drift mobility;
\item PF dependence in weak fields remains unchanged without
respect to trap depth $\Delta$ and concentration $c$;
\item in stronger fields a linear field dependence $\ln\mu\propto E$
emerges.
\end{enumerate}
These conclusions seem to be in striking contradiction with recent
experimental observations for transport of holes in doubly doped polymer layers.
\cite{vj,wbpg,bors_last} Molecules of
one dopant, added in small concentration, and possessing significantly
lower ionization potential, served as traps for charge carriers.
In these studies it was found
that for shallow traps the PF dependence remains mainly
untouched \cite{vj,wbpg}, while for deep
traps a linear field dependence was observed in the
whole field range. \cite{vj}
The most puzzling experimental result is the unusual dependence of the mobility
on trap concentration $c$ \cite{vj,bors_last}
\begin{equation}
\mu \propto 1/c^{n},
\label{cn}
\end{equation}
with $n > 1$ instead of expected
dependence with $n=1$ for trap-controlled transport.
Again, this result does not agree with the
theoretical dependence where $n=1$. \cite{QG_98}

\begin{figure}[htb]
\begin{center}
\includegraphics[width=3in]{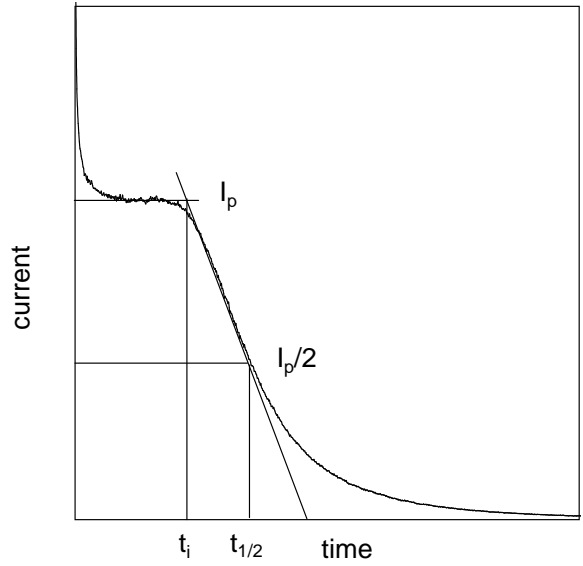}
\end{center}
\caption{Two methods of mobility calculation from experimentally measured
photocurrent transient: 1) $\mu_{i} = L/t_{i}E$
where $t_{i}$ is the time of intersection of asymptotes
to the plateau and trailing edge of the transient;
and 2) $\mu_{1/2} = L/t_{1/2}E$ where $t_{1/2}$ is the time for photocurrent to decay
to the half of its plateau value.}
\label{fig_mu}
\end{figure}

Quite unexpected solution for all these puzzles was suggested in
ref. \citen{image99}. Surprisingly, the problem is not connected
with any fault of the theory or inaccuracy of modern experimental technique
but rather with the interpretation of the experimental data, namely with
the particular way of the mobility calculation from the photocurrent
transient temporal dependence.

\begin{figure}[htb]
\begin{center}
\includegraphics[width=3in]{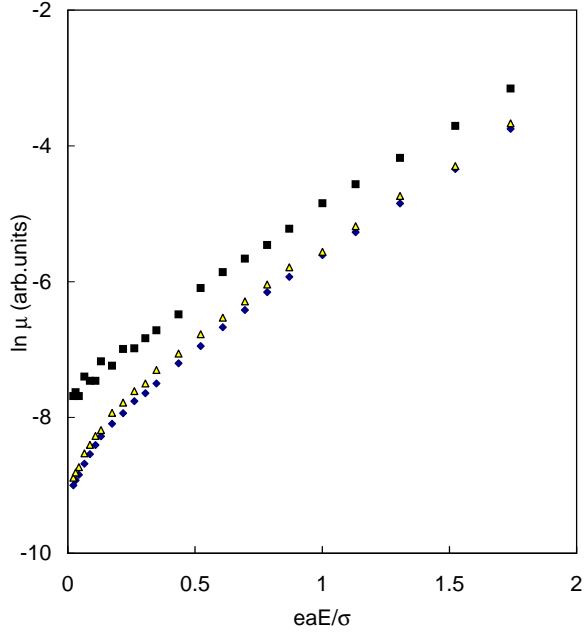}
\end{center}
\caption{Mobility field dependence for dipolar glass with traps
for $\sigma\beta = 3.83$, $\Delta\beta=10$,
$c=0.01$, and $L = 40,000$ lattice planes (equivalent to transport
layer thickness of 30-40 $\mu$m) for different
methods of mobility calculation: $\mu_{v}$ - diamonds, $\mu_{1/2}$
- triangles, and $\mu_{i}$ - squares.
Here $c$ is a fraction of sites occupied by traps.}
\label{fig_x}
\end{figure}

\begin{figure}[htb]
\begin{center}
\includegraphics[width=3in]{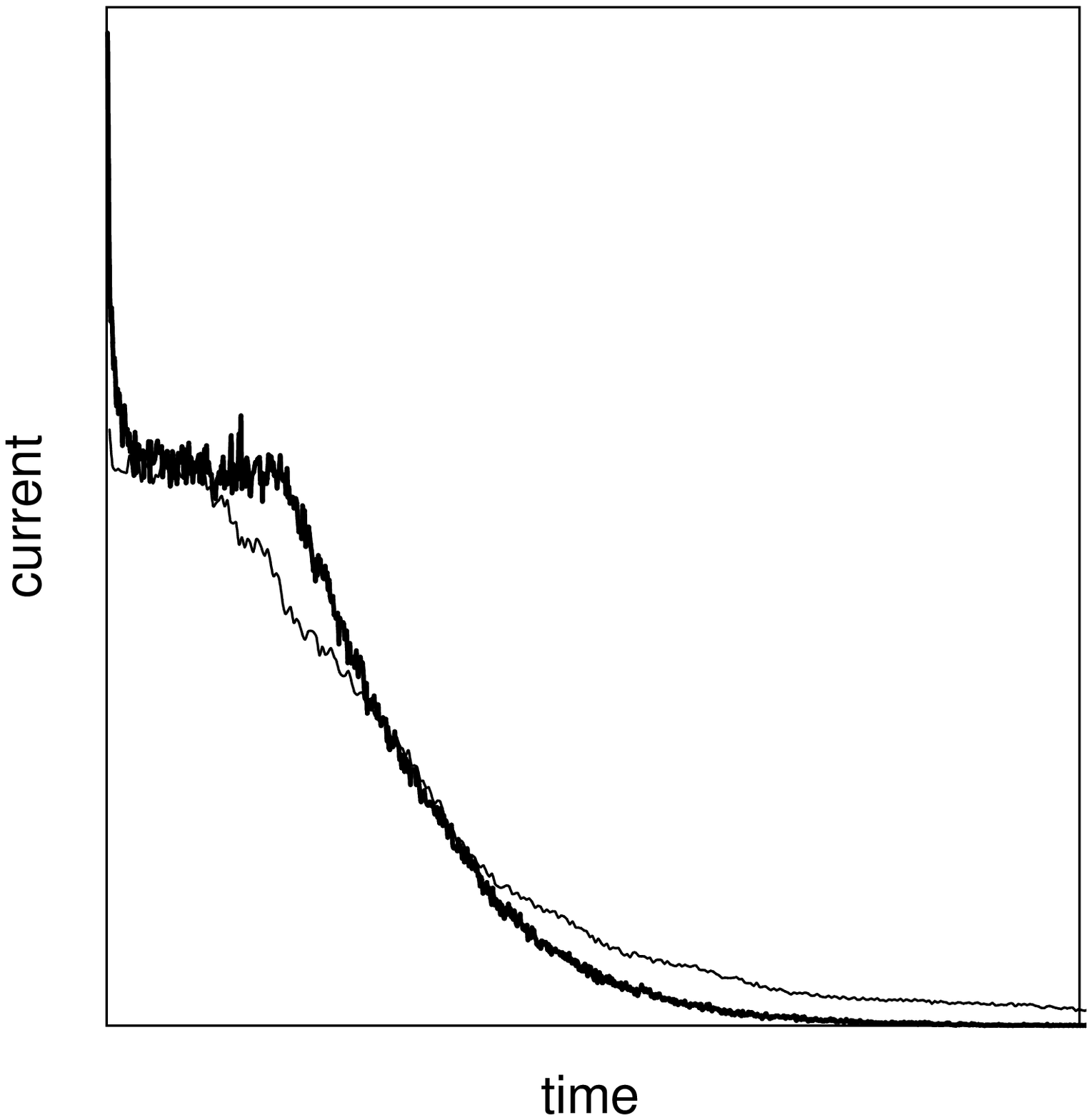}
\end{center}
\caption{Photocurrent transients for dipolar glass with traps for
 two values of $eaE/\sigma$:
0.021 (thin line) and 1.70 (thick line), correspondingly, with time axis
rescaled to make values of $t_{1/2}$ equal for
both transients. All other parameters are the same with Fig. \ref{fig_x}.
Note the different shapes of transients.}
\label{fig_x2}
\end{figure}

In TOF experiment, mobility is usually calculated by two methods (Fig. \ref{fig_mu}).
The first method is the method of choice for most experimental papers.
Unfortunately, this particular method is too sensitive to the
variation of the shape of the transient with variation of $E$ (or $c$)
and tends to overestimate the contribution
of fast under-relaxed carriers.  \cite{image99}
The second method produces mobility that is much more close to the strict definition
of the mobility as $\mu_{v}=\left<v\right>/E$ where $\left<v\right>$ is
an average carrier velocity. The sensitivity of the first method is so significant
that even in the case when good current plateau with $I(t)\approx$ const
is observed in the whole range of variation of the parameter (e.g., $E$),
there is a possibility to obtain principally different mobility
field dependences for $\mu_i(E)$ and $\mu_{1/2}(E)$ (see Fig. \ref{fig_x}).
Note, that for this particular case in the whole field range the
transient is more or less nondispersive and has a well-defined
plateau (Fig. \ref{fig_x2}). We can conclude that in some cases
mobility, calculated by the first method, effectively masquerades
a real field dependence of the true mobility. The same reason
explains an unusual dependence of the mobility on trap
concentration. \cite{image99}

The second method produces much more reliable results and should
be used instead of the first one. This conclusion is in sharp contradiction with
the common belief that field dependences of $\mu_i$ and $\mu_{1/2}$
are essentially the same (experimental evidence for the difference may be found
in literature, \cite{bgm} but it is usually ignored). The use of
$\mu_{1/2}$ is especially important in situations where we have to
choose among not very distinctive alternatives, such as QG or PF mobility
field dependence in nonpolar materials.

\section*{{\sf POOLE-FRENKEL MECHANISM REVISED: \\ EFFECT OF CHARGED TRAPS}}

Original PF mechanism was for a long time considered unacceptable
for explanation of the mobility
field dependence in disordered organic materials due to absence of
charged traps (arguments provided in ref. \citen{spg} suggested
that trap density is certainly less than $10^{13}$ cm$^{-3}$).
Yet in a recent paper the PF model was revitalized by Rackovsky and
Scher, \cite{scher} who argued that a very low density of Coulomb
traps is enough to produce an essentially nondispersive PF charge
transport (they estimated that $10^{11}$-$10^{13}$ ${\rm cm}^{-3}$
should be a sufficient density for transport layer with the
thickness of 10 $\mu$m). Because such low density can easily avoid
detection, they suggested that the PF model still can be
considered as a possible candidate for explanation of charge transport in disordered
organic materials.

Calculation of the PF effect in ref. \citen{scher} was
carried out for the usual case of an isolated Coulomb trap only.
Yet charged traps produce strongly spatially correlated energy landscape,
and the hopping charge motion in such landscape usually has many
features that cannot be captured by the approximation taking
into account interaction of charge carrier with the isolated force
center.

Early indication that charge transport in charged medium may
significantly differ from predictions of the model of isolated
Coulomb trap was presented by Dunlap and Novikov \cite{spie97}.
In recent papers \cite{NIP17,pssb03} this problem has been considered
in more detail for a special case of the medium with equal concentration of
randomly located positive and negative traps
without using the isolated trap approximation. \cite{note1}
The result is in drastic contrast with the approximation of
isolated Coulomb trap.
For example, at $E=E_{\rm crit}=4\pi n \beta e^3/\varepsilon^2$
 (here $n$ is trap density) a transition from mobile to immobile carriers occurs
and average carrier velocity $\left<v\right>=0$ for $E<E_{\rm crit}$ in
the infinite medium,
so in finite transport layers the only possible regime is a dispersive transport
with mobility depending on thickness. \cite{spie97,pssb03}

By accident, mobility field dependence for $E<E_{\rm crit}$ retains
the PF form, but its temperature dependence and, more important,
concentration dependence differ drastically from the prediction of
the isolated trap model. \cite{pssb03} For small concentration of traps
\begin{equation}
\ln\mu \approx -P\left(c,\beta\right)+
2\left[ea\beta \left(E-E_{\rm crit}\right)Q\right]^{1/2},
\label{PF_true}
\end{equation}
\[
P(c,\beta)=\frac{4\pi
h_a^3}{3(\ln c)^2}, \hskip10pt Q(c,\beta)= -\frac{8\pi h_a^4
}{3(\ln c)^3}, \hskip10pt h_a=\frac{e^2\beta}{\varepsilon a}.
\]
Here the lattice version of the model is considered assuming that
fraction of sites occupied by charges is small $c\ll 1$ but
$c\exp h_a \gg 1$.

Experimental test of the mobility dependence on the concentration
of charged traps should be the best test of the result
(\ref{PF_true}). By now there are no experimental data on this
dependence (actually, no experimental data at all on charge transport
in disordered organic materials containing controllable concentration
of static charges; quite possibly, such data are very difficult to
obtain). Nonetheless, the mobility temperature
dependence (\ref{PF_true}) seems to be too strong to describe existing
experimental data.
For this reason we believe that the model of charged traps cannot be suggested
as a serious candidate for explanation of charge transport
in disordered organic materials.

\section*{{\sf CONCLUSIONS}}

All variety of disordered organic materials could be subdivided into
different classes on the basis of their spatial correlation properties in
the distribution of energies of transport sites. Materials
from different classes have different transport properties though
differences are not very pronounced. In order to discriminate between different
classes of the mobility field and temperature dependence,
analysis of experimental data should be carried out with utmost care.
In this respect an early attempt to characterize all materials in
a unified manner (the Gaussian Disorder Model) is unjustified while
we can now understand, why it was so successful.
To some extent charge transport theory is ahead of experiment: major
predictions (mostly concerning charge transport in nonpolar materials)
still lack experimental test.
For example, the reasonable question is: which model,
quadrupolar glass or Los Alamos, is more suitable to describe charge transport
in nonpolar PPVs? A careful and purposeful experimental study
of the transport properties of nonpolar materials could resolve this problem.

\bigskip
\bigskip
\noindent
\textbf{Acknowledgements} I am greatly indebted to A.V. Vannikov,
P.M. Borsenberger, D.H. Dunlap, P.E. Parris, V.M. Kenkre, H. B\"{a}ssler,
A. Kadaschuk, and Z. Soos for numerous discussions.
I am especially indebted to S. Nespurek for providing me
with data on transport properties of polysilanes.
Partial  financial support of the International Science and
Technology Center grant 2207 and the Russian Fund
for Basic Research grant 02-03-33052 is acknowledged.
This work was partially supported by the University of New Mexico and
utilized the UNM-Alliance LosLobos Supercluster at the
Albuquerque High Performance Computing Center.

\newpage


\begin{thebibliography}{20}
\bibitem{note1} The case of charged medium
with unequal concentrations of positive and negative traps is unusual because
in this medium energetic disorder is spatially inhomogeneous
and the nondispersive charge transport is not possible at all.
\end{thebibliography}
\end{document}